\documentclass{article}

\usepackage{arxiv}

\usepackage[utf8]{inputenc} 
\usepackage[T1]{fontenc}    
\usepackage{hyperref}       
\usepackage{url}            
\usepackage{booktabs}       
\usepackage{amsfonts}       
\usepackage{nicefrac}       
\usepackage{microtype}      
\usepackage{lipsum}
\usepackage{graphicx}
\title{The Rise of Technology in Crime Prevention: \\ Opportunities,  Challenges and Practitioners' Perspectives}

\author{
  Dario Ortega Anderez\\
  School of Science and Technology\\
  Nottingham Trent University\\
  Nottingham, UK, NG11 8NS \\
  \texttt{dario.ortegaanderez02@ntu.ac.uk} \\
   \And
 Eiman Kanjo \\
  School of Science and Technology\\
  Nottingham Trent University\\
  Nottingham, UK, NG11 8NS \\
  \texttt{eiman.kanjo@ntu.ac.uk} \\
     \And
 Amna Anwar \\
  School of Science and Technology\\
  Nottingham Trent University\\
  Nottingham, UK, NG11 8NS \\
  \texttt{amna.anwar2016@my.ntu.ac.uk} \\
    \And
  Shane Johnson \\
  Dept of Security and Crime Science\\
  UCL Jill Dando Institute of Security and Crime Science\\
  London, UK, WC1H 9EZ  \\
  \texttt{shane.johnson@ucl.co.uk} \\
  \AND
  David Lucy \\
  Metropolitan Police \\
  London, UK, SW1A 2JL\\
  \texttt{David.M.Lucy@met.police.uk> } \\
}

\begin{document}
\maketitle

\begin{abstract}
Criminal activity is a prevalent issue in contemporary culture and society, with most nations facing unacceptable levels of crime. Technological innovation has been one of the main driving forces leading to the continuous improvement of crime control and crime prevention strategies (e.g. GPS tracking and tagging, video surveillance, etc.). Given this, it is a moral obligation for the research community to consider how the contemporary technological developments (i.e. Internet of Things (IoT), Machine Learning, Edge Computing) might help reduce crime worldwide. In line with this, this paper provides a discussion of how a sample of contemporary hardware and software-based technologies might help further reduce criminal actions. After a thorough analysis of a wide array of technologies and a number of workshops with organisations of interest, we believe that the adoption of novel technologies by vulnerable individuals, victim support organisations and law enforcement can help reduce the occurrence of criminal activity.
\end{abstract}


\section{Introduction}
Criminal activity continues to be a major concern in modern civilisations, with most nations facing unacceptable levels of crime and delinquency \cite{Shaw2003}. Crime is a general term which embodies a wide array of criminal actions. Across offence types, academic literature and official statistics indicate significant variation in both risk and reporting rates. For instance, statistics from the UK Office for National Statistics \cite{Elkin2019} suggest that vehicle-related thefts are amongst the most frequently experienced crimes, whereas homicides are relatively rare. With regards to reporting rates, while these are high for crimes such as domestic burglary, domestic abuse (DA) has been identified as one of the most under-reported offence types. A common assumption is that victims of DA, especially women who are abused by their partners, are reluctant to report domestic violence (DV) to the authorities \cite{Felson2002,Hague2005,Herzberger2019}.

While crime in its multiple forms has generally declined over the last two decades \cite{farrell2014crime}, there are notable exceptions. For instance, as per UK 2019 figures, violence against the person has shown a relevant increase as compared to the previous year \cite{Elkin2019}. Increases in specific types of crime are often attributable to changes in opportunity. For instance, increases in online activity have led to significant opportunities for cyber-enabled (made easier by the internet, such as fraud) or cyber-dependent (only possible because of the internet, such ransomware attacks) crime.  
For other criminal offences, such as domestic abuse, while figures may be increasing, part of the reason for the trends observed in the UK Police data is that the gap between what is experienced and what is reported is closing \cite{Stripe2020}. That is, improvements on police recording practices and the support services provided appear to be encouraging victims to come forward more than they once did. This is critical as crime cannot be prevented without a clear understanding of the scale of a specific offence and of how these are committed. Put differently, the challenges associated with addressing crime are not limited to reducing offending. Providing (and signposting) support services for those affected, and capturing the true scale of the problem can play a vital role in preventing further criminal activity.   

\begin{figure}
    \centering
    \includegraphics[width=0.85\textwidth]{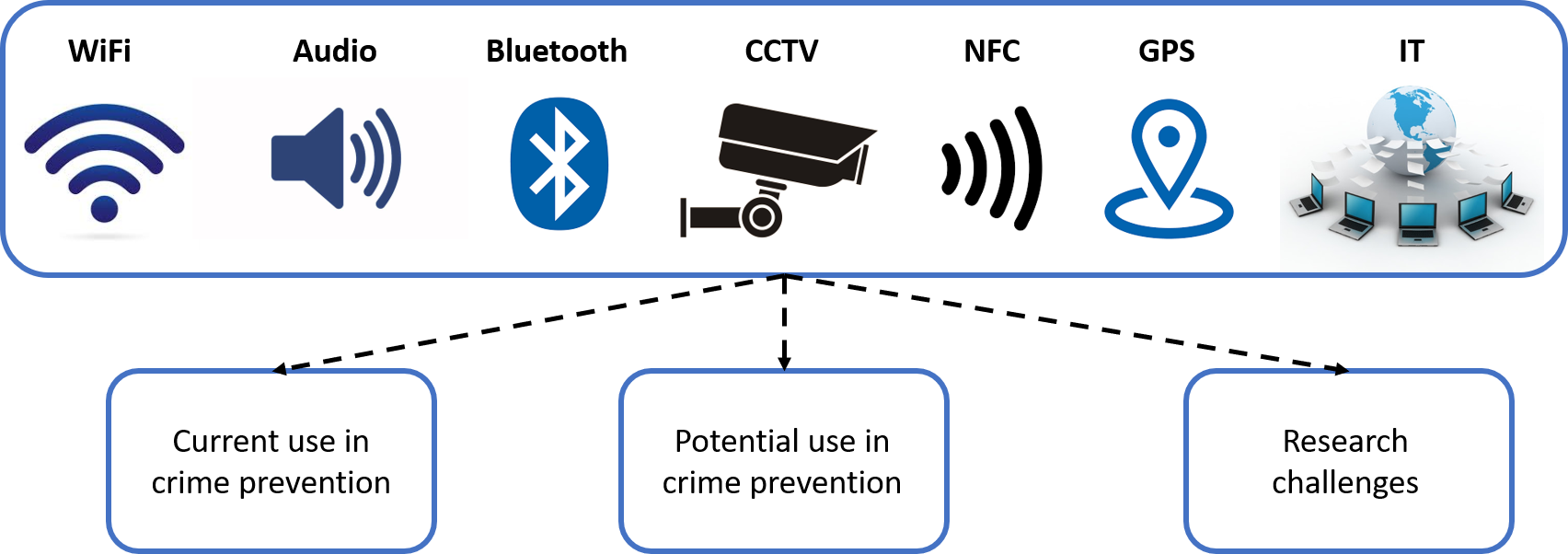}
    \caption{Paper outline with the main technologies discussed.}
    \label{fig:outline}
\end{figure}

As alluded to above, technological innovation can generate opportunities for crime. However, technology has also been one of the driving forces that has enabled the improvement of approaches to crime prevention, and poling more generally \cite{Chan2001}. For example, the first police technology revolution, which incorporated the use of two-way radio, automobiles and telephones, facilitated substantial change in the way police organisations operate \cite{Harris2007}. As several authors suggest \cite{Chan2001,Stroshine2005,Harris2007}, it is now time for the second revolution, with such a revolution implicating once again a change in police administration and organisation. The use of Information Communication Technology (ICT) has and is likely to continue to be capital in terms of understanding crime problems and responding to them. However, to date, compared to industry, in policing the exploitation of developing technologies has been somewhat limited in the crime reduction enterprise. 

This paper provides a review of the different technology-driven solutions that have been employed for the purposes of crime prevention in the last few decades, as well as a critical analysis of how contemporary hardware and software-based technologies, such as ubiquitous sensors and machine learning, could be used for improving the reporting and the prevention of criminal activities in the (near) future. In what follows, the technical advantages and disadvantages of a wide array of sensing devices are discussed in the context of crime prevention (see Fig. \ref{fig:outline}), with the aim of identifying new research directions and encouraging research activity in the field. 

\begin{figure}
    \centering
    \includegraphics[width=0.40\textwidth]{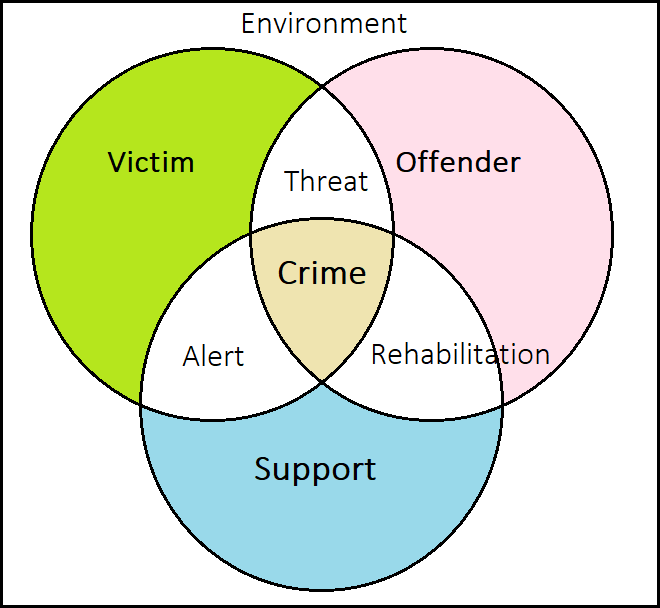}
    \caption{The relationship between crime stakeholder and the environment.}
    \label{fig:crime}
\end{figure}

In addition to the analysis of opportunities new technologies present to monitor and prevent crimes, a number of sandpits and co-design workshops with victim support, policing and end users organisations were conducted. These discussions have also helped in understanding the relationship between the different crime stakeholders (see Fig. \ref{fig:crime}), as well as how technology can not only support victims but also prevent or reduce crime levels by monitoring offenders behaviour and their interaction with the surrounding environment.

The remainder of this paper is organised as follows (see also Fig. \ref{fig:outline}). Section \ref{sec:previous work} reviews current applications of pervasive computing technologies used in crime prevention. Section \ref{sec:ChallandOpport} discusses the challenges and opportunities identified for the employment of pervasive technologies to prevent, detect and report criminal activity. Section \ref{sec:Challenges} presents the main challenges for the adoption of novel technology-driven crime prevention tools. Section \ref{sec:conclusions} concludes the manuscript with a summary of the main findings and provides recommendations for future work on the developments of technological tools for crime prevention. 

\section{Technology-driven Solutions for Crime Prevention}\label{sec:previous work}
The advent of pervasive computing and the IoT provides new potential opportunities for technology-driven solutions for crime prevention \cite{Chen2015,Byrne2011}. This section discusses the working principle of such technologies as well as (where appropriate) how these are currently used for preventing crime. The section is organised according to the main source of information or sensing device used by each of the discussed technologies.

\subsection{Camera-based Technologies}\label{subsubsec:camera}

In recent years, Closed Circuit Television (CCTV) surveillance has emerged globally as a mainstream crime prevention measure \cite{Piza2019}. CCTV technology is already present in many public places such as railway stations, airports, office buildings and on the street. Moreover, as early as 2005, estimates suggest that in the United Kingdom there were already over 4 million cameras in place – an approximate ratio of 1 camera per 14 citizens \cite{Norris2005}.

While CCTV cameras were formerly employed as a means to report crime, support police officers with prosecution processes or as a supporting evidence in court of justice, the increasing developments in computer vision and machine learning provide opportunities for investigating CCTV surveillance as a crime prevention mechanism \cite{Piza2019}.

CCTV cameras are typically intended to reduce crime through several mechanisms. The first is to deter offenders who fear being identified committing a crime on camera.  The second is to provide law enforcement agencies with an opportunity to identify crimes in progress or those that are about to happen, to provide them with an opportunity to intervene.  Where these two mechanisms fail, CCTV cameras can record evidence that a crime occurred, who was involved and what took place, which can be used in criminal prosecutions. The first mechanism relies on offenders perceiving that CCTV is effective, otherwise there will be no perceived risk.  The second relies on staff monitoring (possibly many) camera feeds and being able to identify suspicious activity.  The third relies on the camera(s) being able to capture an event in progress, which may require it to pan and tilt accordingly. It is not hard to imagine why failures associated with human error might emerge, given the difficulties associated with these tasks.  However, developments in computer vision and machine learning, which are leading to a continuous improvement in applications such as object and face recognition \cite{Klontz2013,Abdullah2017} or gait analysis\cite{Bouchrika2016,Zeng2016}, clearly provide opportunities for CCTV surveillance that might enhance its effectiveness \cite{Piza2019}.

\begin{figure}[t]
    \centering
    \includegraphics[width=0.50\textwidth]{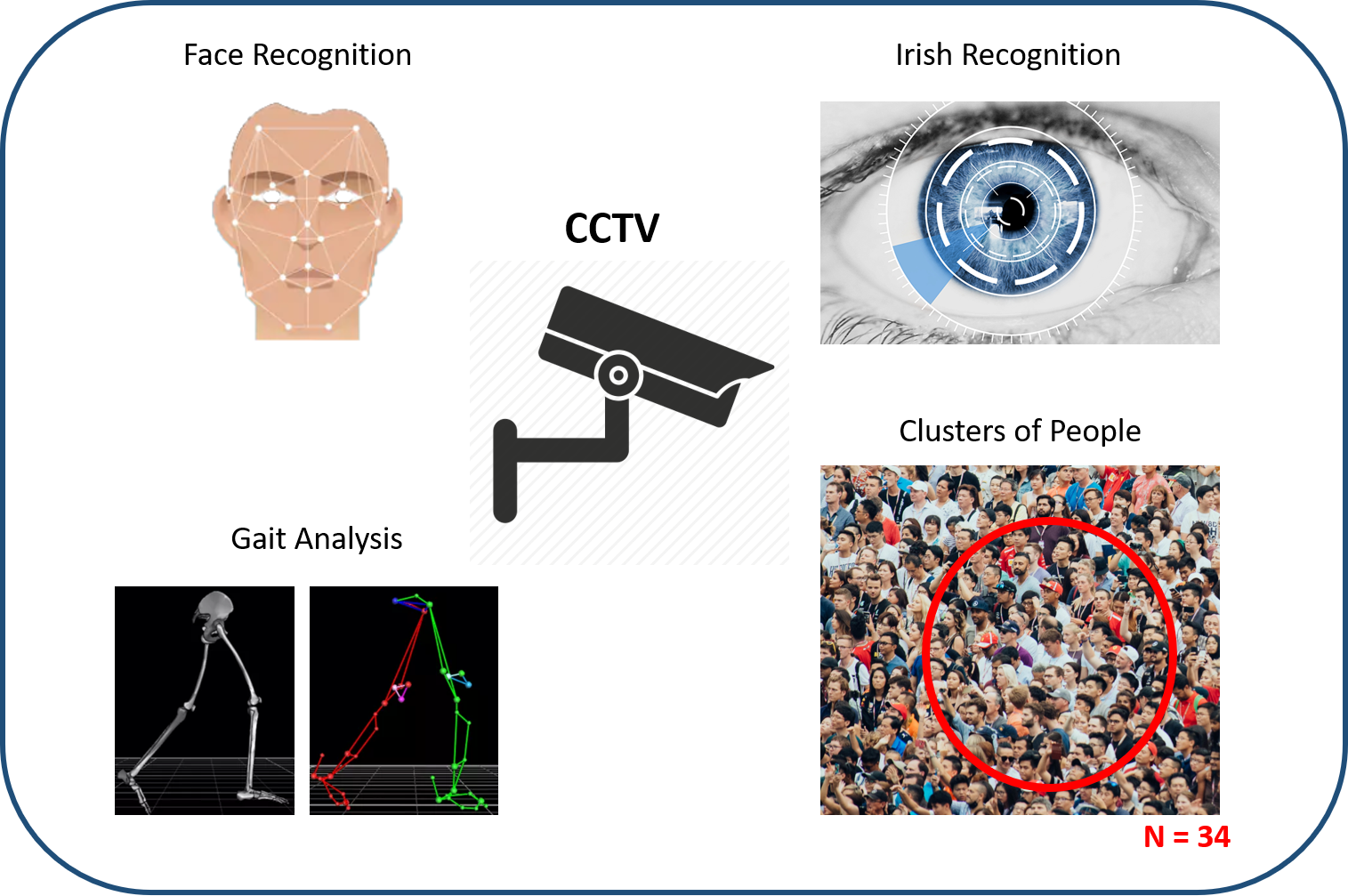}
    \caption{Applications of CCTV surveillance on crime prevention}
    \label{fig:CCTV}
\end{figure}

CCTV surveillance systems are constantly being upgraded to incorporate the latest soft technology features \cite{Byrne2011}. These include on-the-edge feature extraction and machine learning classification algorithms. As a result, it is now possible to extract different patterns and personal bio-metrics from video sequences, which can be a posteriori used for person identification. In this sense, CCTV technology has been investigated in different applications (see Fig. \ref{fig:CCTV}). These include the automatic detection of suspicious anomalies such us unattended bags in mass transit areas or crowded venues \cite{Bhargava2007}, iris recognition-based security systems which deny access to buildings to unauthorised personnel \cite{Gragnaniello2015}, intrusion detection systems (IDS) in unauthorised areas employing motion tracking techniques \cite{Ketcham2014,Kim2018}, automatic robbery detection in banks via object and human posture detection \cite{Kakadiya2019}, and even as a means of crowd detection and congestion analysis for safety purposes \cite{Li2014}.

\subsection{Electronic Monitoring and Global Positioning System-based Technologies}\label{subsec:EM}
\begin{figure}[b]
    \centering
    \includegraphics[width=0.50\textwidth]{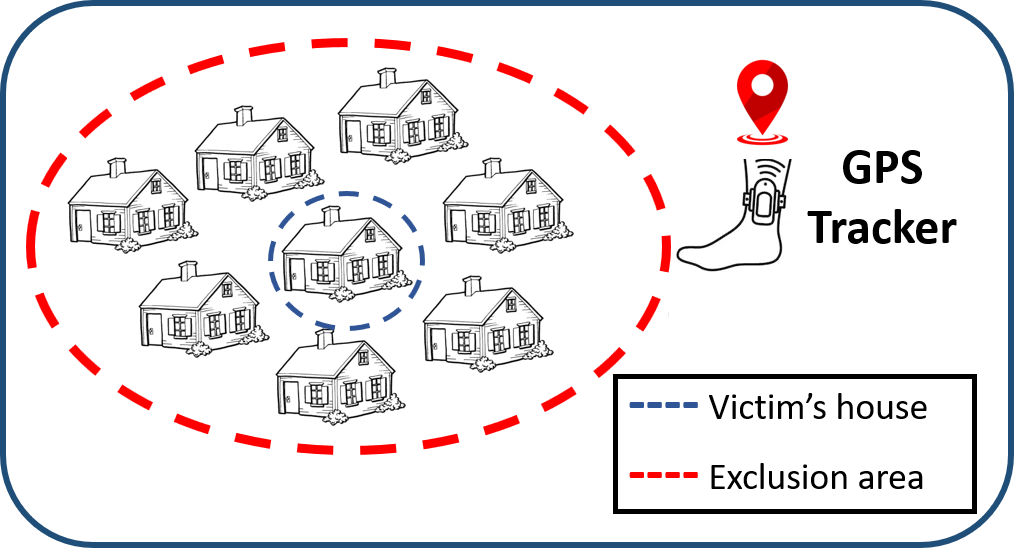}
    \caption{GPS Geo-location .}
    \label{fig:GPS}
\end{figure}

Electronic Monitoring (EM) began in the early 1980's in the US and spread rapidly after positive initial claims in reducing control deficits in community supervision \cite{Burrell2008}. First EM systems employed radio frequency identification (RFID) technology. RFID technology is based on a tag with a unique identifier which sends data to an electronic reader through wireless radio frequency waves, enabling its identification and tracking. RFID technology was first employed to confine offenders (or pre-trial defendants) to a particular location (usually their home) \cite{Brayford2013}. The work in \cite{Finn2002} showed that placement of sex offenders on EM programs reduced the likelihood of and postponed their return to prison. The use of RFID technology was then extended to the protection of victims of domestic violence. Victims were provided with receivers so that they were alerted when an offender was present within a pre-established control perimeter, which was normally set to be around the victim's home \cite{Erez2006} (see also Fig. \ref{fig:GPS}).
     
The second generation of EM technologies incorporated the use of Global Positioning System or GPS. GPS is a satellite-based global navigation system which provides geo-location through the use of a network of satellites orbiting the Earth at an altitude of approximately 20,200 km. To estimate the geo-location, a GPS receiver intercepts the signals of at least three network satellites at regular intervals of time. A posteriori, based on the time it takes to receive each of the satellite signals, the geo-location of the GPS receiver is calculated via trilateration. The use of GPS technology as a crime prevention tool has gained increasing attention since late 1990's \cite{Grommon2017}. The ability to customise exclusion zones and provide instant alerts if these are violated, has extended the use of electronic monitoring to sex offenders and post-work release offenders \cite{Downing2006}. A pictorial example of how GPS technology is used in this context can be seen in Fig. \ref{fig:GPS}. Further applications, such as the tracking of terrorist suspects to gain insights into their spatial and temporal behaviour, have recently been proposed \cite{Griffiths2017}.

Overall, as outlined by \cite{Belur2020}, EM is an effective tool towards preventing recidivism, specially for sex offenders.

\subsection{Short-Range Wireless Communication Technologies}\label{subsec:Bluetooth}
Short-range wireless communication transceivers such as Bluetooth and WIFI are widely incorporated into many portable and mobile devices, including laptops, mobile phones and smart-watches. During the manufacturing process, a wireless module is assigned a unique identification (ID) in the form of a 48-bit Medium Access Control (MAC) address. This address is then used to identify and authenticate a device when communicating with other wireless devices. For example, Bluetooth (BT) devices can interact with other nearby BT devices within their signal range ($10m$ to $100m$, depending on the radio transceiver) by sending and receiving radio waves within a band of 79 different frequencies centred at 2.45 GHz. Using such radio waves, along with the identification capabilities provided by the unique MAC address assigned, a Bluetooth device can continuously monitor other Bluetooth devices nearby (within its signal range) and also identify the type of device associated with such MAC address (i.e. whether it is a smart-phone or a laptop for example). In addition, the Received Signal Strength Indicator (RSSI) provides an estimated measure of the power present in a received radio signal. As shown by previous RF-based research \cite{Benkic2008,Adewumi2013,Awad2007}, RSSI can then be used to estimate the approximate distance a Bluetooth receiver is from a Bluetooth emitter. 

\begin{figure}
    \centering
    \includegraphics[width=0.50\textwidth]{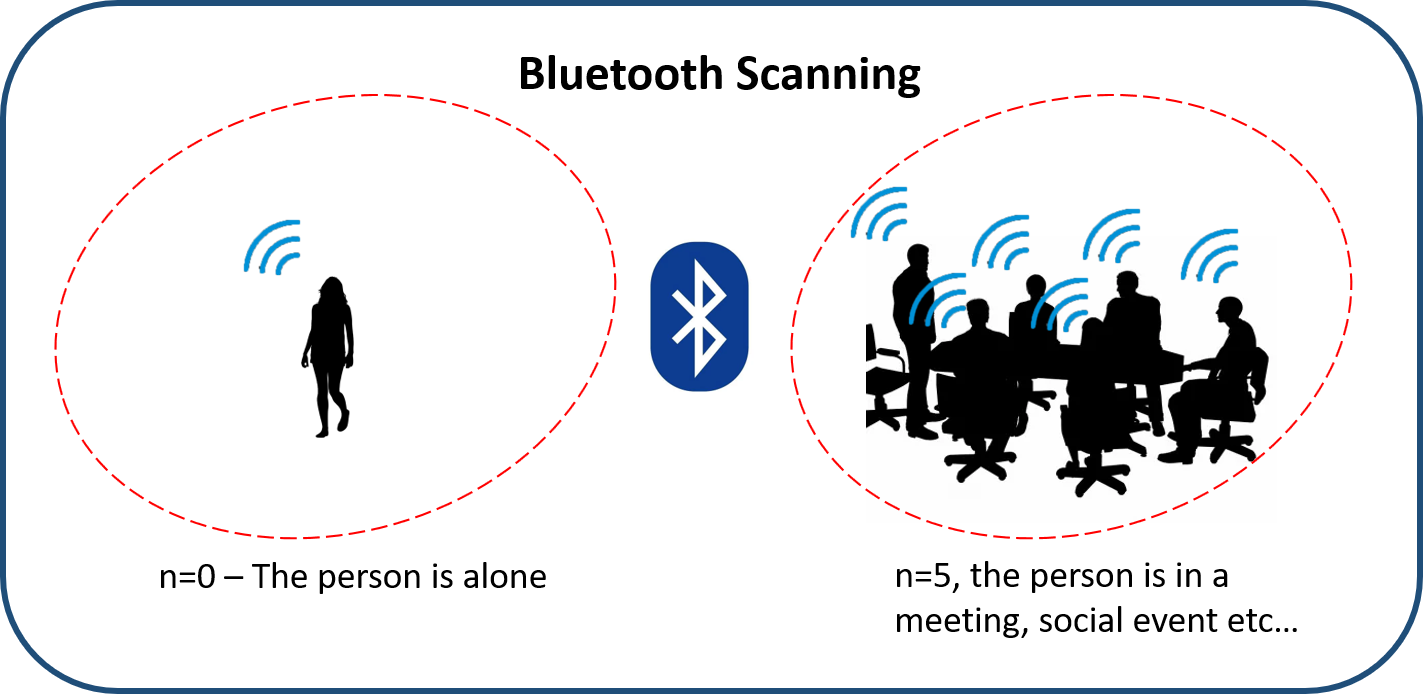}
    \caption{Social context monitoring via Bluetooth device scanning.}
    \label{fig:bluetooth}
\end{figure}

Exploiting the above characteristics of this technology, various researchers have employed Bluetooth technology to estimate the surrounding social context of a person \cite{ONeill2006} (see Fig. \ref{fig:bluetooth}) or to estimate pedestrian flows at specific places \cite{Nicolai2007,Nishide2013}. The authors in \cite{Chen2015} discuss how such monitoring capabilities could be used to prevent or investigate the kidnapping of infants and elementary school children.  They note that kidnappings tend to take place when children are alone, and that a system could be used to continuously log the Bluetooth devices near to a child and, when none are detected, an accelerometer is used to monitor their activity. Ultimately, they suggest, a mobile application (where the Bluetooth logs can be visualised) could be provided to the child's parents so that they can monitor their children's social context throughout the day.   

\subsection{Audio-based Technologies}\label{subsec:audio}
Audio-based technologies have been used for many years to anticipate criminal actions. Common examples include the interception of calls by police officers (wire tapping) or the use of recording devices as evidence collection mechanisms \cite{Fishman1976}. With the current advances in audio processing and machine learning techniques, whereby several parameters and bio-metrics can be automatically computed from raw audio recordings, the scope of audio-based technologies have experienced a great expansion. For instance, gun shot detection audio-based technologies have been recently evaluated in the US \cite{Lawrence2018}. Gun shot detection is a novel technology which employs a network of microphones, typically installed in high crime areas, which discriminates gunshots from other types of noises and computes the spatial coordinates of the location where the shot was fired.
Besides, the presence of Intelligent Virtual Assistants (IVAs) or Chatbots such as Amazon Alexa or Google Assistant is becoming increasingly popular \cite{Chung2017}. Such IVAs incorporate speech recognition capabilities allowing users for asking questions and making requests to different interfaces. In addition to speech recognition, it is now possible to count the number of speakers in a conversation via speaker diarization \cite{Anguera2012}, to infer the sentiment (mood) of individuals by analysing their voice \cite{Bird2018} or yet recognise a speaker by his/her voice with considerably low error rates \cite{Richardson2015,Snyder2018}. These advances clearly provide opportunities for implementing audio-based crime prevention tools.

\subsection{Affective Computing}
Recent research has realised that there is a significant relationship between the physical health and emotional state of an individual \cite{Picard2016}. Given this, the field of affective computing is receiving increasing attention in the last few years \cite{Sano2013,Kanjo2015,Can2019}. Affective computing or emotional intelligence is the study and development of systems for the recognition, processing and interpretation of human affects. Typically, this is performed with the use of wearable devices or smart textiles by which various physiological signals related to stress levels are measured, processed and interpreted. Electro-dermal activity (EDA) and heart rate variability (HRV) are two major examples of physiological signals employed in affective computing. Although affective computing is still as its infancy, several studies have shown that such signals can be translated into relevant features which ultimately lead to estimations of human stress levels. For instance, the research studies in \cite{Ollander2016,Liu2018} have made use of the MIT Stress Recognition in Automobile Drivers Database \cite{Healey2005} to classify between three different stress levels (low, medium and high) in three different driving scenarios.          

Although, to the best of our knowledge, affective computing has not been employed in the context of crime prevention, the anticipation of high levels of stress could assist the prevention of crime. As suggested by literature in the field \cite{colvin2002coercion}, mental disorder and violence may each be rooted in the stress levels at which an individual lives. On another note, mental disorder has been shown to increase the chances of committing crime. For instance, in the study conducted by \cite{Hodgins1992} -using a Swedish birth cohort selected at random- it was found that men with major mental disorders were more than ten times more likely than men with no mental disorders to have criminal offense records and four times more likely to be registered for a violent offense. Other work \cite{Coker2000,Jones2001,Stewart2017} suggest that, on average, victims of domestic violence have poorer mental health, which can lead to further issues including depression or anxiety. Although as aforementioned, sentiment analysis is still at its infancy, this technology clearly shows the potential to be explored as a crime prevention mechanism.

\subsection{Information-based Technologies}
In addition to the technologies discussed above, a wide array of information-based crime prevention solutions have been developed in the last years. 

One of such technologies is Crime Mapping (CM), also known as hot-spot policing \cite{Hutt2018}. CM refers to the process of conducting spatial analysis to map, visualise and analyse crime patterns\cite{Santos2016}. This allows for the identification of crime hot spots in conjunction with other crime trends and patterns\cite{Abbas2017,Johnson2010}. Such information can then be used to optimise the location of human or/and technological resources. As various works suggest, the identification of hot spots is an effective software-based technology for optimising the use of resources and ultimately prevent crime \cite{Braga2017,Williams2017}.

Risk assessment is another key information-based technology for crime prevention. Risk assessment is used to assess the risk of recommitting crime by offenders under correctional control. According to the survey conducted in \cite{Byrne2009}, a majority of serious crimes are committed by a small fraction of people during the first months of probation parole. Risk assessment tools make use of predictive models to identify such subgroup of people so that appropriate surveillance/supervision is granted to those cases. Likewise, information technology is used to identify the likelihood of a terrorist attack or a serious violent event occurring at certain places, including schools, airports or train stations among others \cite{Byrne2011}.   
Another application where information technology has been adopted to prevent crime is in the development of computer software to track individuals' interactions on various social media sites \cite{Soghoian2011}. The monitoring of such suspect's interactions is then used to identify abnormal behaviours which can potentially be related to crime intentions.

\subsection{Crime Prevention Mobile Apps}\label{sec:mobileapps}
The number of downloads of mobile apps has been steadily growing worldwide for the past decade. Along with this, the average time per day spent by adults on their smartphones is growing at a fast pace and this is forecasted to continue. For instance, in the UK, that figure is expected to go over 4 hours in 2021 \cite{Fisher2019}. With this in mind, the development of mobile apps as a means of crime prevention seems highly beneficial for promoting and enhancing social safety. In this context, a wide array of applications have been proposed to help either with the prevention of crime or with the reporting of crime that has already occurred. Apps for crime prevention can be broadly divided into two categories, namely apps to be adopted by policing institutions and those directed to the general public. The remainder of this section presents a wide array of mobile apps for crime prevention with a special focus on those related to the reporting of emergency situations.

\begin{figure}
    \centering
    \includegraphics[width=0.60\textwidth,height=0.50\textheight]{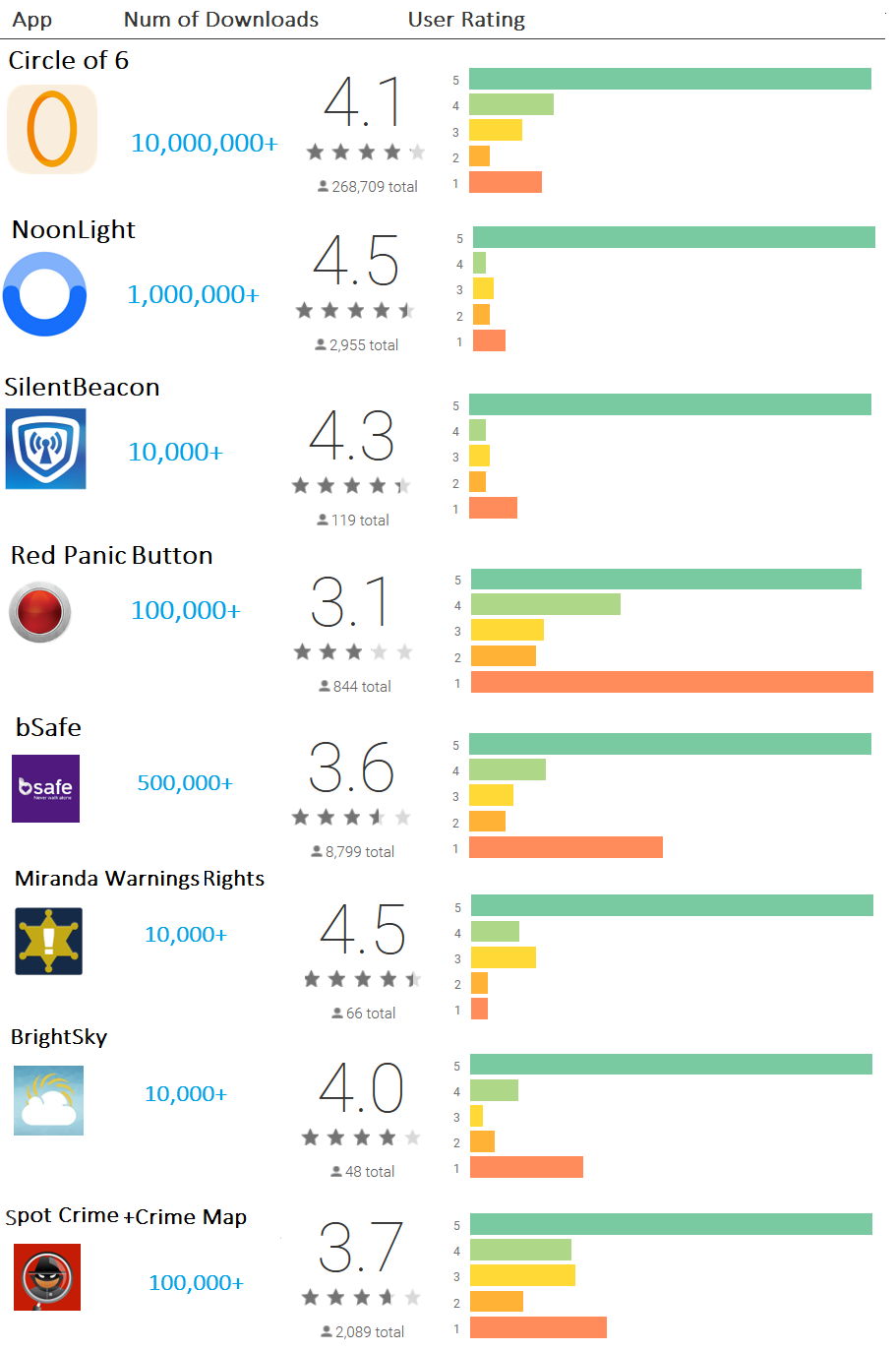}
    \caption{Summary of indicative popular crime prevention apps in the Google Play store}
    \label{fig:apps}
\end{figure}

\textbf{Mobile Apps for Emergency Situations:} Numerous mobile apps have been developed in recent years to facilitate the communication of emergency and panic situations to close relatives and policing institutions. For instance, Circle of 6 \cite{Circleof6} is a mobile app that enables its users to quickly contact a user-defined list of six people and share their current location along with an emergency message. 
In a similar way, Noonlight \cite{Noonlight} provides a mechanism to alert emergency institutions by pushing and holding a button embedded in the app. When the button is released, the user is asked for her/his pin number. A posteriori, the user just needs to enter the pin to indicate he or she is safe. If the pin is not provided, the situation is immediately reported to emergency institutions along with the live GPS geo-location of the user. A similar working principle to that of Circle of 6 and Noonlight is provided by Silent Beacon \cite{Silentbeacon}. In this case, a portable device in the form of a panic button is provided alongside the app. Once the button is triggered, the device communicates the action to the smartphone app which then automatically calls 911 while communicating the emergency situation and the location to a user-defined list of contacts. A similar functionality is offered by Red Panic Button \cite{RedPanicButton}, which in addition to the above features, incorporates the automatic publication of emergency messages on Twitter. Another example is bSafe \cite{bSafe}, which allows a list of user-selected contacts to track the user's way home using GPS information. In addition, as with the apps presented above, it allows the communication of emergency situations while also collecting evidence in the form of audio and video recordings.\\   
Fig. \ref{fig:apps} lists all the apps reviewed in this section including their app icon, number of downloads and user ratings.  

\textbf{Mobile Apps (miscellaneous):}
In addition to the apps to aid in the reporting of emergency situations, a number of further mobile apps which likewise aim at the prevention of crime in varying ways are available for the public. One of such mobile apps is BrightSky \cite{BrightSky} (10k+ downloads on GooglePlay, scores 4.0). BrightSky is a mobile app mainly directed towards victims or potential victims of domestic abuse, being the first smartphone app to provide a UK-wide directory of specialist domestic abuse support services.
BrightSky enables users to locate their nearest support centre by searching their area, postcode or current location. Other features include a tool to log incidents of abuse in text, visual and audio forms. In addition, a number of questionnaires to assess the safety of a relationship as well as further information and resources regarding domestic abuse, sexual consent, harassment and stalking are provided withing the app.  
SpotCrime+ Crime Map \cite{SpotCrime} is a mobile app which provides a visual platform in the form of map with different crime icons to keep the public aware of the crimes that occur in their area in real time by retrieving data from a US crime database. With this information, the public can both avoid getting close to areas of high crime and provide any evidence or statement concerning the crime of interest in their area. A similar approach is followed by the app Police Crime Statistics Map \cite{PoliceCrime}, which provides a similar platform to that of SpotCrime but for UK citizens.
In addition to the above, various apps have also been developed to assist police officers in a wide array of their numerous work duties. An example of these is Miranda Warnings Rights \cite{Miranda} which provides an easy to access reference to assist police officers in mirandizing suspects. 

From the analysis of the above, it can be concluded that smartphone apps provide effective mechanisms to both victims and police organisations to prevent crime in different ways. Further testing, development and promotion of crime prevention apps can be crucial to reduce criminal activities under different circumstances.

\subsection{Edge Computing and on Device Processing}\label{sec:edge}
Traditionally, the majority of the processing for data intensive applications was done on a central cloud to take advantage of fast and powerful computing infrastructure. However, major issues concerning latency, security and privacy can be identified in the use of cloud-based systems for crime prevention. 

In this regard, a new trend of processing the data on the edge is emerging. The motivation behind edge computing is that of performing the data processing as near as possible to the point of data production. With this, the privacy and latency issues present in cloud-based systems can be significantly mitigated. AI is gradually finding its way into embedding systems which are becoming smaller and less power demanding, while offering fast processing power and low latency at an increasingly attractive cost.  

\begin{figure}
    \centering
    \includegraphics[width=0.50\textwidth,height=0.25\textheight]{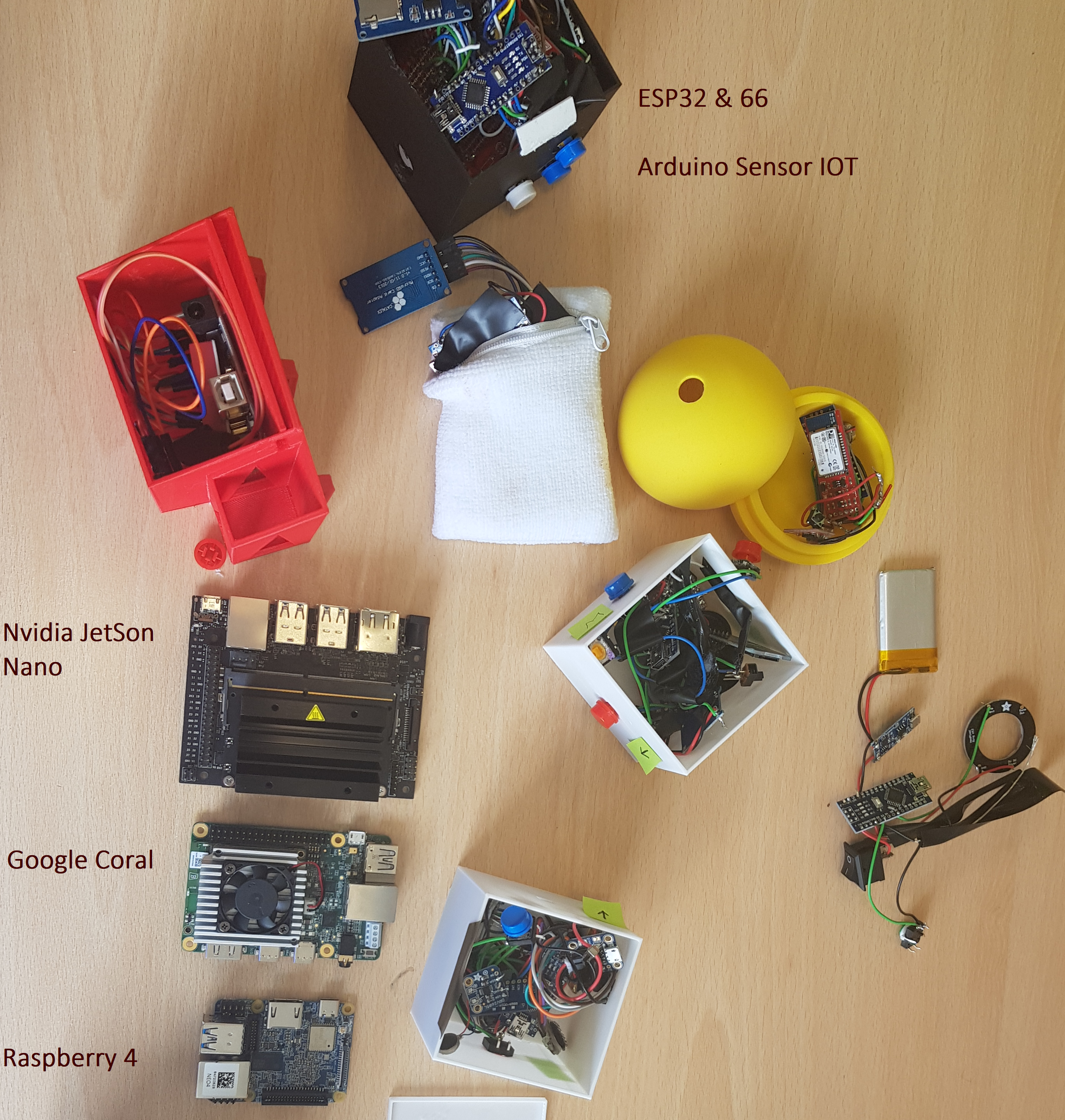}
    \caption{A number of modern edge devices with potential use in behaviour analysis for crime detection and prevention.}
    \label{fig:edge}
\end{figure}

A number of off-the-shelf edge computing devices suitable to carry out heavy signal processing and machine learning applications are already available (see Fig. \ref{fig:edge}). For instance, both Nvidia and Google have recently released their respective development boards, namely Jetson Nano and Google Edge TPU, with the aim of enabling users to develop and run AI applications on the edge. In addition to their portability and the privacy advantages they offer, such boards are supported by sophisticated development kits that consist of a SOM (System-on-Module) connected to a development board which incorporate numerous connectors like USB and Ethernet to share the data gathered when desired. Furthermore, the above devices also support major deep learning frameworks and tools such as TensorFlow.

The above, alongside current developments in the field \cite{Suarez2019}, suggest that edge computing is finding its way into city centers for the early detection and prevention of criminal actions.  

\section{Opportunities}\label{sec:ChallandOpport}
This section discusses the potential use of the technologies presented in Section \ref{sec:previous work} for the prevention of criminal actions. A summary of the information discussed in this section is provided in Table \ref{tab:technical_comp}. 

\begin{table*}[]
\caption{A list of technology-driven solutions for the prevention of criminal actions.}
\centering
\resizebox{\columnwidth}{!}{%
\renewcommand{\arraystretch}{1.5}
\begin{tabular}{|l|l|l|l|}
\hline
\textbf{System}            & \textbf{Technical Advantages}    & \textbf{Technical Disadvantages} & \textbf{Potential Applications}                                                                 \\ \hline 
CCTV-based  
& Person identification & Privacy concerns at home environments  & Collection of evidence                                                                      \\
                  &  Activity Recognition   & Visibility             &  Detection of unattended objects                                                             \\
                   & Object recognition      & Lighting conditions   &    Robbery prevention                                                             \\
                  &                         & Occlusion               & Physical violence prevention                                                              \\
                  &                         & Limited field of view    & Deterring offenders                                                              \\
                  &                         & High power consumption    &                                                             \\ \hline \hline
Bluetooth-based & Ubiquity                & Limited functionality on its own & Social context analysis / Detection of clusters of people           \\ 
                  & Social context analysis &    MAC Randomisation  & Stalker detection                                                                                  \\
                  & Low energy consumption  &    RSSI Instability       &   Home and vehicle robbery prevention                                                                         \\ \hline \hline
GPS-based         & Ubiquity             & Data unavailability      & Outdoors abuser tracking                                             \\ \hline \hline
NFC-based        & Ubiquity            & Activation by user & Emergency alert mechanism \\
                  & Reduced size        & & Easy mechanism for accessing information
\\ \hline \hline
Audio-based       & Sentiment Analysis      & Lower degree of detail as compared to CCTV & Evidence collection \\
                  & Ubiquity                &  & Detection of violent actions (i.e. glass breaking)                                                                   \\
                  & Spherical sound field &                 &                                                                       \\ \hline \hline
Information-based & Behavioural Analysis     & Long-term conclusions & Abnormality detection in social media                                                                  \\
                  &                         & User-dependent            &  Hot-spot identification                                                            \\
                  &                         & Data-dependent & Crime prediction \\ \hline                                                                       
\end{tabular}
\label{tab:technical_comp}
}
\end{table*}

CCTV surveillance has been widely employed to prevent and report crime under different circumstances. The continuous developments seen in pervasive computing as well as in computer vision, allow for the on-board computation of different bio-metrics (i.e. face recognition or gait analysis) and for the extraction of patterns which are key to identify suspects as well as activities taking place within its field of view. Despite the great potential shown by CCTV surveillance on preventing various type of crimes in public and outdoor places \cite{Piza2019}, several drawbacks are found on its application inside home environments (i.e. abusive relationships). First, after the installation, the device is typically visible. While this is paramount for applications where the aim is to deter crime, where the aim is to collect evidence surreptitiously it would be better to conceal the device. Second, a home CCTV surveillance system would require the installation of numerous CCTV cameras, since the view field is limited to the room at where the camera is mounted. Third, CCTV cameras are sensitive to lighting conditions and occlusion (\textit{i.e.} aggressive behaviour in the night could not be detected). Ultimately, as outlined by numerous works, there is a common reluctance towards the adoption of video cameras in home environments due to privacy concerns \cite{Anderez2020}. As a summary, video cameras alongside the latest machine learning and pattern matching algorithms can be of great use to avoid crimes in outdoor settings as well as in private businesses and public transport but alternative sensing technologies are preferred for home environments.       

Short-range communication technologies such as WiFi and Bluetooth have been proven useful for analysing the social context of a person \cite{ONeill2006,Chen2015}. For instance, by continuously logging the presence of Bluetooth devices around a device and their respective Received Signal Strength Indicators (RSSIs), it is possible to know the alone and accompanied periods of an individual. In addition, it is possible to infer the social context of a person by the analysis of the variation in the number of surrounding devices across time. For instance, when a person is walking on the street, the social context could be inferred by the changes on the number of devices detected across time. Likewise, when a person is alone, the number of surrounding devices is not expected to vary. Furthermore, the closeness of an external device to a person could be evaluated by the long-term analysis of the corresponding logged MAC address of such external devices across time. This could be of great use to identify stalking scenarios, and more importantly, the periods of time when the potential stalker and the potential victim are found alone. However, an increasing number of devices make use of MAC randomisation. Therefore, to achieve such aim, hard settings to elude MAC randomisation should be employed. This could be achieved by providing pre-programmed hardware such as wearable devices through which a unique identifier is broadcasted. With this, a novel electronic monitoring system with a similar working principle to that of the contact tracing utilised to prevent the spread of the COVID-19 could be deployed. In cases where offenders are known to the police and have to use EM as a condition of their sentence, this system could be employed to complement current systems which generally use GPS technology, providing altogether a more robust system which would overcome the frequent lack of signal shown by GPS technology in indoor environments. Current portable devices such as smart phones or smart watches already incorporate Bluetooth transceivers, allowing for the continuous log of surrounding devices. In addition, current Bluetooth transceivers make use of a low energy protocol, namely Bluetooth Low Energy (BLE), reducing the power consumption required to operate. A further potential application with the use of Bluetooth device scanning is the prevention of robberies at home environments. To do so, Bluetooth Beacons could be installed at home environments to detect surrounding devices which remain within the Bluetooth antenna range for suspiciously long periods of time or/and at a suspicious time, raising an alarm when appropriate. Furthermore, short-range communication technologies could be employed for the monitoring of different areas through the detection of intrusive scenarios.

GPS technology can be of great use to ensure security distances are met. Reported cases of DV in which the offender is forced to wear a GPS tracking device are a good example. However, despite the fact that most portable devices incorporate GPS technology, personal geo-locations are only shared with services the users opts to share the data with, therefore being not publicly available. Given this, GPS technology can be considered of little use for unreported cases of crime. On the contrary, short-range communication technologies constantly broadcast a signal to the surrounding devices, thus overcoming the drawback shown by GPS technology in this context. 

Short-range tagging technologies such as Near-field communication (NFC) technology are gaining increasing attention and made available in  most of the latest generations of smart phones. NFC technology is a set of communication protocols for short-distance communications (4cm or nearer) between two electronic devices, namely a NFC reader and a NFC tag. Such communication is achieved through the NFC reader sending a signal to a pre-programmed battery-free NFC tag which sends back its function or application to the reader. Common examples of the use of NFC technology include contact-less payment and door card readers. Given the reduced size of NFC tags, these could be potentially embedded in a wide range of everyday objects such as clothes, key rings or furniture to be used as an alerting mechanism in cases of emergency. Likewise, tag tapping could be used to trigger evidence collection mechanisms such as microphones or video cameras in a quick, easy and efficient way. NFC technology would be therefore a great complement to the current emergency reporting and evidence collection mobile apps presented in Section \ref{sec:mobileapps}, which will enable the triggering of the different apps' built-in actions on the quiet.           

Audio-based technologies have been employed for many years as a crime prevention tool through the use of wire tapping and recording devices \cite{Fishman1976}. In addition, novel applications such as gunshots \cite{Ratcliffe2019} and screaming/shouting detection \cite{Laffitte2019} are surging with the advent of machine learning and pervasive computing. From the psychological perspective, as compared to CCTV surveillance, audio monitoring is considered as a less invasive way of monitoring individuals \cite{Chen2005}. This therefore indicates the privacy concerns found in the use of video surveillance systems in home environments are mitigated by the use of audio sensing devices, making them an attractive mechanism for the collection of evidence. In addition, as discussed in Section \ref{subsec:audio}, the current advances in audio processing and machine learning allow for the computation of various elements which can be crucial to prevent domestic violence. Let's consider a violent argument between a potential victim of domestic abuse and her/his abuser. Through audio diarization, their voices can be separated, therefore allowing for the analysis of each of the audio streams separately. Bad language within the conversation could be identified through the use of speech recognition. In addition, the mood (i.e. anxious, violent, sad) of each person could be estimated through the use of audio sentiment analysis techniques. Overall, audio-based technologies can be of great use as a means of evidence collection. Further advances in audio and natural language processing would grant this technology with crime prevention capabilities by the identification of emergency situations inferred from sound.  

Information-based crime prevention systems make use of various data sources to predict the risk of a crime being committed. For instance, as previous work shows \cite{Matthews2017}, there is a common tendency from abusers to isolate victims of domestic abuse by controlling and monitoring their electronic devices and social media, with several reported cases of account hijacking, or even destroyed devices. Given this, information-based technology could be employed to identify deviations from the regular personal social media behavioural pattern from a potential victim of domestic abuse, therefore providing clues for the detection of further criminal actions. Likewise, analysing the social media behaviour of individuals, it is possible to employ probabilistic models to estimate the probability for a specific person to commit crime. In addition, crime mapping techniques \cite{Eck2005} which allow the identification of crime hot spots could be employed to optimise the use of personal and technological resources. Although the information gained from the above techniques could benefit the identification of potential criminal actions, various important limitations are found. First, the use of trend analysis techniques may require long-term studies. That is, the behaviour of individuals may need to be monitored for some time to establish a useful profile of activity. This can imply a long delay between the occurrence of the criminal actions and the conclusions drawn from the analysis. Second, in the context of personal social media behavioral analysis, a previous active use of social media by subjects of interest may be required. Third, may people seek to remain anonymous online by using pseudonyms \cite{Van2015}. Where they do, it will be difficult to develop profiles of their activity that could be used for the purposes of crime prevention. To use such data, it may be necessary to have an individual consent. Fourth, even if deviations from personal patterns are identified, care would need to be taken in interpreting these as it cannot be assumed these are surely caused by the experience of abuse or by the willingness to commit crime. Used in the right circumstances though, such approaches may have value.

\subsection{Feedback from End User Organisations and Practitioners}
The opportunities new technologies present to monitor and prevent domestic abuse were explored during a series of workshops with academic experts, practitioners, and a selection of engineers and technologists, including representatives from police forces, government, charities, trusts, voluntary support groups and end users. 
Few workshops took place between June 2018 and May 2020. The content of the different workshops is summarised as follows:
\begin{enumerate}
    \item Exploratory Workshops (World Cafes): We held two workshops to explore how developing technologies might help to reduce the risk of domestic abuse. The aim of the first workshop was to work with domain experts to produce a "requirements brief” that mapped out a handful of priority problems, together with any factors that constrain how they are dealt with now and how they might be dealt with in the future. The aim of the second workshop was to identify possible solutions to the problems identified. A mixture of domain experts, designers and engineers were invited to the second workshop.
   \item Co-Design Workshop: Existing examples of crime prevention technologies were discussed to explore the opportunities they offer. Also, the difficulties that survivors might face in calling for help or recording evidence discreetly were discussed. Few potential technologies that might help in supporting both victims, police and aid evidence gathering were introduced. Participants particularly liked the portability of small Edge Computing platforms including wearable devices and the different methods of interaction as compared to smartphones. Participants were excited by the concept of proximity detection techniques being able to sense the presence of individuals and any abnormal activities in the surrounding environment and the nearby vicinity as many vulnerable individuals might find it difficult to record such incidents. The possibility for IoT devices to provide evidence and a reliable journal was also intriguing as the participants had not acknowledged these technologies before.
   \item Co-Creation Workshop: A reduced number of possible technological solutions including tagging and proximity detection were discussed along with hand-on demonstrations. First the potential use of short range tagging technology as part of a panic alarm was discussed. The scenario proposed was one where small tags “dots” are placed in home environments, or in public places, which then can be used to send an alarm or as an information access point. Proximity detection technologies \cite{woodward2020digitalppe,anderez2020covid} were also discussed due to their potential in ensuring regular offenders keep their legal distance from victims. Fig. \ref{fig:proximity} shows how the alerts can be sent out depending on the proximity of the devices of interest. An alternative suggestion considered the use of internet connected devices, such as smart speakers, to automatically detect incidents of abuse whilst these were in progress. To reduce false alarms, it was suggested that such a system could scan for multiple signals to include slammed doors, raised voices, particular keywords in verbal exchanges, or other indicators of threats of risk. We discussed the practicality of such a system and how to protect user’s privacy, as well as how triggering actions could be tailored to the individuals in question to increase the sensitivity of the system. Once triggered, data can be stored locally or on the cloud, and alerts can be sent to trusted third parties.
   \begin{figure}
    \centering
    \includegraphics[width=0.60\textwidth,height=0.20\textheight]{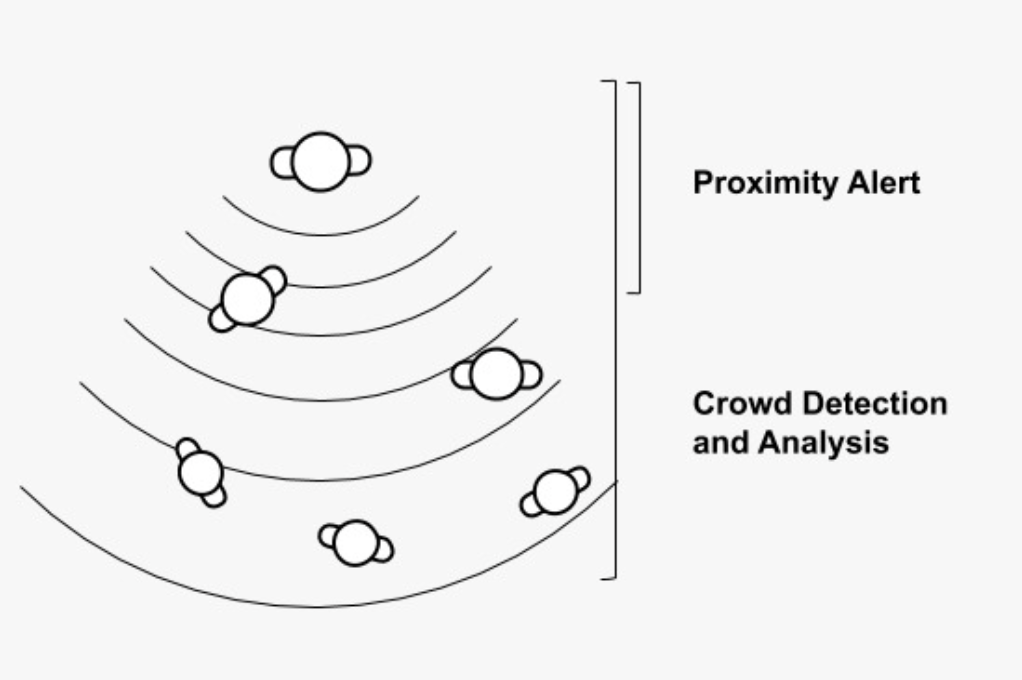}
    \caption{ Illustration of proximity detection technology around crowds.}
    \label{fig:proximity}
\end{figure}

  \item One to one discussions with End Users Organisations: Based on the discussions in the previous workshops, the aim of these series of online discussions was to narrow down the possible development routes and to come up with a list of paper prototypes which were discussed with each organisation to understand their implications and limitations.
  
  \item Design Evaluation Workshop: In the last of these workshops, the previously discussed technologies narrowed down the proposed prototypes, including the use of short-range communication mediums for alert and recording of abnormal activities in close proximity and also the use of AI to detect violent behaviour. During the workshop the discussion converged on the benefits of a single edge device aimed at crime prevention and the collection of evidence that would protect users privacy. The use of edge computing would mean that data would not need to be stored remotely and the available processing power would mean that computational intense processing (e.g. voice and sentiment detection) could be carried out on such a device. Personalised and more adaptable interfaces or functionalities of the device then can be developed targeting diverse groups of people.

\end{enumerate}

The series of co-design and co-creation workshops carried out concluded that there is a need for a range of technological solutions to address crime prevention, as a one-size-fits-all solution could not feasibly address the different issues and potential users. In addition, a single organisation cannot solve all the problems alone, a collaborative effort is thus required to reach optimal solutions. 
One of the workshops participants David Lucy from the Met Police explains:

\textit{ "The Police Service is facing increased and changing demand, putting pressure on front line services. Powerful new technologies provide a way for the Police Service to prevent crime and serve their communities more efficiently. Crime is changing in the digital age. The police service must continue to adapt, to deal with new and emerging cyber-crimes, whilst still addressing the demand from things such as acquisitive crime and domestic abuse. Demand from both emerging and existing crimes can be reduced through the use of new technology. We have been engaged in a series of co-design sessions and sandpits with collaborators from Nottingham Trent University (NTU) and the University College London (UCL) discussing the potential of smart devices and technologies to prevent crimes and support vulnerable individuals. By engaging closely in these activities, we have learnt about the potential of proximity detection and tagging technologies in alerting relevant support organisation or family members to an imminent risk. We firmly believe these should be further explored and trialed since they show the potential to support policing with preventing crime, supporting investigations but ultimately achieving the best outcomes in the pursuit of justice and in the support of victims".}
The workshops have also helped in identifying and understanding the key research challenges faced by researchers who work in this research field. Such challenges are summarised in the section below.

\section{Research Challenges}\label{sec:Challenges}
\begin{figure}
    \centering
    \includegraphics[width=0.65\textwidth]{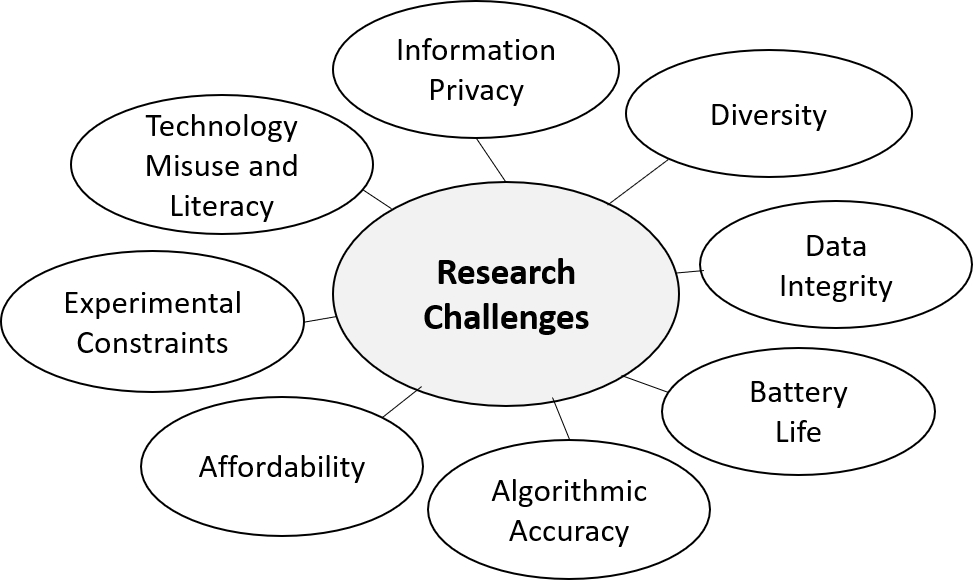}
    \caption{Challenges of adopting technological innovations for Crime prevention. }
    \label{fig:challenges}
\end{figure}

This section presents the main research challenges faced at the development of pervasive computing technologies for preventing criminal activity (see Fig. \ref{fig:challenges}). 
\subsection{Information Privacy}
There has been much research conducted in the recent years in relation to the privacy implications of smart devices. However, as stated in \cite{Habibzadeh2019}, system developers are normally compelled to meet strict deadlines to avoid losing competitive advantage, leading to immature products which fail to satisfy the expected privacy and security requirements of their specific target applications \cite{Khatoun2017}. 

A large number of users may get understandably uncomfortable with sending their personal data to remote data stores or clouds which they cannot see or control. Transferring user data over networks (including secure networks) make systems vulnerable to theft and distortion. To be perceived as legitimate by the public, and to meet ethical and legal standards, privacy requirements and concerns need to be satisfactorily addressed.

The above privacy concerns can often be addressed by guaranteeing that a user’s personal data will be processed at the point of collection and will not be transferred remotely. By employing edge computing and on-device processing techniques, chipsets can potentially be exploited at or close to the data source for performing computational tasks instead of transmitting all the data and processing it in a central server. A competent trade-off between personal privacy and computational power should certainly be optimised for each crime prevention tool if this is to be widely employed by the public. 

\subsection{Diversity and Scalability}
Developing AI technologies to detect explicit personal behaviours or actions requires the collection of data from a large number of users, to enable predictive models incorporate sufficient variability in order to generalise well on unseen data. Therefore, it is crucial to carry out data collection processes on a large number of experimental participants, as well as to consider the differences that may exist between different groups of individuals. For instance, let's imagine one is planning to develop and launch a novel audio-based system to recognise violent behaviours for the prevention of domestic abuse in households within the UK. The first problem a system developer may encounter is that not every household speaks English. A speech recognition model which is only trained with data from native speakers only, would certainly not perform well with data processed from non-native speakers. Additionally, there are other sound characteristics, such as tone, phonetics, intonation and melody which may vary considerably between different languages or even between different accents of the same language. Thus, although it may involve tedious processes of data collection, it is crucial to identify the target group and incorporate an adequate level of inter and intra-group variability when developing intelligent systems. This issue becomes even more crucial when dealing with sensitive matters like crime where false negatives can translate into serious personal health-related consequences.     

\subsection{Data Integrity}\label{subsec: Data Integrity}
Data integrity refers to the consistency, accuracy and completeness over the entire life cycle of the different data retrieval, data transmission, data storage and data processing units of a system. Maintaining data integrity is crucial for various reasons. First, data integrity ensures recover-ability and trace-ability. Second, the decision-making made by intelligent systems is driven by the data it is developed upon, thus the accuracy of such systems strongly depends on the integrity of the data they are developed, maintained and functioned with.      

\subsection{Battery Life}
Energy reduction and sustainability have become major issues in the technical and social agendas in the last years. The ubiquitous and pervasive computing research communities are systematically facing a strain between usability and sustainability. On the one hand, users express an increasing interest in purchasing more sustainable products. On the other hand, they do not wish to do so at the sacrifice of their comfort.

The widespread use of mobile and portable devices such as smart-phones or tablets comes alongside an ever increasing demand for mobile apps, including resource demanding applications such as face, speech or activity recognition. Although resources like memory, bandwidth and processing power are constantly being improved to keep up with the increasing demand for additional storage, communication and processing power, battery capacity grows only at an approximate rate of 10\% per year \cite{Orsini2016}. 
However, increasing battery capacity is not the only way to improve battery lifetime. Alternatively, research efforts have been and should still be made to explore ways to reduce power consumption from mobile devices.          

\subsection{Accuracy and Experimental Constraints}
Intuitive optimisation of various parameters often result in better and more accurate models, with error rates being continuously lowered. However, there exist various limitations and challenges which do not normally allow Artificial Intelligence (AI) applications to exhibit error-free performances. First, machine learning and deep learning algorithms typically require large amounts of data to successfully undergo the training phase. In addition, as mentioned in Section \ref{subsec: Data Integrity}, such data should incorporate the adequate inter and intra-subject variability for the classification or regression models to be able to generalise well on unseen data samples. This means that in addition to the need to collect large amounts of data, that data has to be variable enough to adequately represent the characteristics of the target population and avoid large generalisation or out-of-sample errors. Further to ensuring the collection of "enough" data samples, adequate feature engineering and machine learning techniques can be a crucial factor to optimise the accuracy achieved by AI-based crime prevention systems. A research challenge therefore arises from the need to improve the performance achieved by the state-of-the-art in the corresponding fields. For instance, as a current survey on sentiment analysis indicates \cite{Shoumy2020}, the classification accuracy achieved by the state-of-the-art on sentiment analysis using audio recordings, is in the range of 72.9\% to 85.1\%. This means that if an audio-based verbal abuse system was to be developed, it would be wrong 14.9\% of the cases.         
The behavioural data required for these models (e.g. \cite{Bird2019} for video Processing \cite{Kanjo2019} for emotion recognition \cite{Ortega2019,Anderez2020} for activity recognition), need to be collected under highly variant free-living scenarios rather than a controlled settings. 

\subsection{Affordability}
Cost was a key factor discussed during the focus group as end users and other supporting organisations would need the device to be as inexpensive as possible if it was to become adopted into practice. In this regard, while tagging technologies relatively cheap, applications requiring high processing power may depend upon expensive hardware. 

Edge computing enables processing all or part of the data at the location it is collected. Data that is only of ephemeral importance can be crunched on the edge device itself. This is in contrast to cloud-based systems, where data is sent to large, remote data centers for processing. In accordance with Moore’s Law \cite{Gustafson2011}, small devices at the edge have become more computationally powerful. If the trend is to continue, it is only inevitable that switching to edge platforms would offer much affordable solutions in the long run. This territorial proximity to the endpoint is good for both latency and efficiency as it saves networks from unnecessary congestion as well as from carrying sensitive personal data.

The ongoing minimisation and mass production of electronics has enabled a reduction in the cost of edge computing devices, whereby various complex computations (such as AI and data processing) can take place on-board. However, the more data and computing intensive an application is, the more data storage and processing power are required, with this having an impact on the price to be paid by end users. In this regard, the adoption of high processing power technologies by the public has two main research challenges associated to it. First, to keep up with the ongoing miniaturisation and cost reduction of electronic components. And secondly, the optimisation of signal processing and machine learning algorithms so these can be adopted utilising a lower computing power.     

\subsection{Technology Misuse and Technology Literacy}
Advancement in technologies enables law enforcement and voluntary services to support victims and reduce or prevent crimes. However, the increasing complexity and communication capabilities present in these technologies have also opened new pathways for data interference\cite{Freed2018}.

There is little empirical research published concerning the use of technology in intimate partner stalking, as most of the current efforts are focused on online abuse on social media or texting \cite{Delanie2017}. Within these, the work in \cite{Delanie2017} conducted a survey with 152 domestic violence advocates and 46 victims. The study found that modern technologies can potentially give perpetrators multiple tools to control and manipulate people and that technology-facilitated stalking needs to be treated as a  serious offense. There is thus a need for non-judgemental responses from service providers and law enforcement to victims experiencing such abuse. As practitioners observed, advising victims to switch off devices, to withdraw from social media, or to change their profile or telephone numbers, is putting an enormous burden of responsibility on the victim to adjust their behaviour \cite{Woodlock2020}. Additionally, disengagement from technology can mean that victims are increasingly uncontactable, which can impact the type and the timing of the support they receive from services.

There is a great need to increase public awareness of the use of spyware to commit abuse and stalking. Likewise, law enforcement and victim support and rehabilitation organisations will benefit from learning about the latest development of technologies that might help vulnerable individuals or prevent technology misuse.

\section{Conclusions and Directions for Future Work}\label{sec:conclusions}
Different technology-driven solutions and the potential adoption of contemporary smart pervasive, machine intelligence systems and miniature technologies for crime prevention have been considered and evaluated.
To our knowledge, this is the first paper that reviews recent technological innovation in crime prevention.

As discussed above, in our view, there is enormous potential associated with the adoption of short-range communication technologies as a tool for crime prevention. These technologies can be employed alongside current approaches, such as GPS monitoring, to provide a more robust proximity detection system, able to detect proximity in both indoor and outdoor environments. 

Smart and short-range tags could be employed to provide an instant access to information and also for emergency reporting. As discussed, several apps to report emergency situations and collect evidence of those situations have been developed in recent years. However, these apps commonly require the user to open the app and perform an specific action.    

Furthermore, violence detection scenarios in home or work environments, the use of a audio-based technologies appear to be preferred against that of CCTV cameras given the ubiquity, spherical field and lower privacy concerns exhibited by the former systems. As exposed, the use of audio signal processing along with machine learning techniques, allowing for applications such as speaker diarization, speech recognition, person identification and sentiment analysis, could be key to identify violent language as well as violent actions. 

From a technical viewpoint, standalone systems employing a single sensing technology exhibit distinct limitations. However, the combination of technologies, can be of great use to identify violent scenarios, which can lead to the prevention of further occurrences and therefore to the ultimate prevention of criminal activity.    
In conclusion, not all circumstances or situations are the same and there will be no “one size fits all” solution(s). As such, it is important to explore the heterogeneity in offenders, victims, contexts of offending, and offending patterns to understand which solutions might work for whom, and under what circumstances. Beside privacy, scalability, affordability, miniaturisation and personalisation are some of the important factors that need to be considered when designing technologies for crime prevention.  

Nonetheless, future work will embody the conduct of focus groups with co-design and co-creation elements where the findings and conclusions drawn in this paper will be further discussed and analysed with end user organisations and various groups of interest. With this, we aim to gain more insights into the potential, limitations and drawbacks of the discussed technologies and obtain critical advise from experts in the field. The development and implementation of a number of prototypes of the selected crime prevention systems will follow.

\section{Acknowledgement}\label{sec:acknowledgement}

This research was funded by the Dawes Center for Future Crime at UCL. We would like to thank all the participants in our workshops for contributing to the discussion and co-design sessions.
We would like to thank all the participants in our workshops who have openly shared their views and ideas. Each of their individual contributions has helped us to better understand the facets of co-design for crime prevention. The discussions in the workshops have played a pivotal role in the development of our views on this topic.

\bibliographystyle{unsrt}  
\bibliography{template}  







\end{document}